\documentclass[conference]{IEEEtran}
\IEEEoverridecommandlockouts
\usepackage{cite}
\usepackage{amsmath,amssymb,amsfonts}
\usepackage{algorithmic}
\usepackage{graphicx}
\usepackage{textcomp}
\usepackage{xcolor}
\def\BibTeX{{\rm B\kern-.05em{\sc i\kern-.025em b}\kern-.08em
    T\kern-.1667em\lower.7ex\hbox{E}\kern-.125emX}}

\usepackage{fancyhdr}

\begin{document}

\title{Team Learning-Based Resource Allocation for Open Radio Access Network (O-RAN)\\
{%\footnotesize \textsuperscript{*}Note: Sub-titles are not captured in Xplore andshould not be used
}
}

\author{\IEEEauthorblockN{Han Zhang, Hao Zhou, and Melike Erol-Kantarci, \IEEEmembership{Senior Member, IEEE}}
\IEEEauthorblockA{\textit{School of Electrical Engineering and Computer Science,}
\textit{University of Ottawa}\\
Emails:\{hzhan363, hzhou098, melike.erolkantarci\}@uottawa.ca}}

\maketitle
\thispagestyle{fancy}   
\fancyhead{}                
\lhead{Accepted by 2022 IEEE International Conference on Communications (ICC) , \copyright2021 IEEE}
\cfoot{}
\renewcommand{\headrulewidth}{0pt}      

\begin{abstract}
Recently, the concept of open radio access network (O-RAN) has been proposed, which aims to adopt intelligence and openness in the next generation radio access networks (RAN). It provides standardized interfaces and the ability to host network applications from third-party vendors by x-applications (xAPPs), which enables higher flexibility for network management. However, this may lead to conflicts in network function implementations, especially when these functions are implemented by different vendors. In this paper, we aim to mitigate the conflicts between xAPPs for near-real-time (near-RT) radio intelligent controller (RIC) of O-RAN. In particular, we propose a team learning algorithm to enhance the performance of the network by increasing cooperation between xAPPs. We compare the team learning approach with independent deep Q-learning where network functions individually optimize resources. Our simulations show that team learning has better network performance under various user mobility and traffic loads. With 6 Mbps traffic load and 20 m/s user movement speed, team learning achieves 8\% higher throughput and 64.8\% lower PDR.
\end{abstract}


\section{Introduction}

The demand for mobile communications has been continuously growing.
With the emergence of new applications, e.g., virtual reality and massive machine type communications, the data traffic volume has reached an unprecedented level. In addition, these new applications require better quality of service with agility, intelligence and flexibility. In order to meet the increasing demands while keeping capital and operational costs low, architectural reformation is needed to adopt openness and intelligence in the radio access network (RAN). Therefore, the recent concept of open radio access network (O-RAN) is proposed in \cite{b1}. In addition to O-RAN there might be other dissaggregated, virtualized, multi-vendor RAN architectures in the future.   

The O-RAN architecture is designed in a layered and modular fashion, which makes network service more flexible and cost effective. The concept of x-application (xAPP) refers to the network control and optimization applications provided by third-party users. In O-RAN architecture, the near real-time (near-RT) radio intelligent controller (RIC) serves as a safe and reliable platform for xAPPs, providing standardized interfaces and hardware support to ensure compatibility. However, in practice, placing multiple xAPPs, with overlapping objectives, into the same RAN may lead to conflicts. Sometimes these conflicts might be subtle and hard to detect as the xAPPs are very likely to be developed by different vendors. There are mainly three kinds of conflicts between these xAPPs, namely direct conflicts, indirect conflicts and implicit conflicts\cite{b2}. 
For instance, a power allocation xAPP may allocate a high transmission power to one resource block. But at the same time, a radio resource allocation xAPP may allocate this resource block to a user with a small traffic load. These conflicts will waste the scarce bandwidth resource and increase power consumption, thus it is critical to handle such conflicts.

To this end, we propose a team learning based algorithm to eliminate conflicts between xAPPs in O-RAN. In particular, we define a team learning method to simultaneously handle the power and radio resource allocation xAPPs. Note that, our technique is not limited to O-RAN, the developed team learning principles also apply to  future multi-vendor RAN architectures. 
Several prior works have studied the joint optimization of power allocation and radio resource allocation. \cite{b11} proposed a federated learning algorithm to jointly optimize transmit power and radio resource allocation for vehicular networks. In \cite{b4}, a reinforcement learning based method is proposed for joint power and radio resource allocation in 5G. In \cite{b16}, a joint power allocation and radio resource allocation algorithm is proposed for multi-user beam forming. These existing studies assume a traditional RAN architecture where both power allocation and radio resource allocation is done by the algorithms developed by a single vendor. In O-RAN, or other future RAN architectures, network functions of multiple vendors might be used together.  

Our work is different than existing works by a more realistic O-RAN architecture and a novel team learning technique that can be used in any multi-vendor environment. In the remaining parts of the paper, we will refer to O-RAN only as an existing example of a multi-vendor RAN. Firstly, in O-RAN, different network resources, such as power and radio resources, are generally managed by different xAPPs of various vendors. Compared with the existing joint resource allocation scheme, the xAPP based architecture enables higher flexibility for network management. Secondly, we propose a novel team learning based algorithm. 
Team learning is applicable when a team of agents are in the same environment and share part of the observational information \cite{b15}. They learn and choose actions in a distributed manner and cooperate for the same team goal. Compared with most existing multi-agent learning algorithms, members in team learning have a higher degree of independence. %Team members may have heterogeneous architectures and should be seen as black boxes. 
%During the learning process, members should cooperate while maintaining as much independence and data privacy as possible. As a result, some existing multi-intelligence learning are not practicable for team learning, such as centralized training and parameter sharing.

In this work, we consider different xAPPs as members of a team with the same goal of maximizing the throughput of the system. Firstly, we designed two xAPPs, namely power allocation and radio resource allocation, as two agents and both of them use deep Q-network (DQN) to make decisions. Then, we assume agents can exchange information with each other, and they  make decisions based on the environment and other agents' intention. By information exchanging, agents can cooperate better as a team and avoid conflicts over the control of the network. Finally, experimental results show that using our team learning algorithm, we can achieve better network performance. Our proposed team learning method presents 8.8\% higher throughput and 64.8\% lower packet drop rate under 6 Mbps traffic with a user moving speed at 20 m/s.

The rest of this paper is organized as follows. Section \ref{s2} introduces %some 
related works, and Section \ref{s3} defines system model. Section \ref{s4} analyzes the main architecture of team learning based xAPPs interaction. The proposed team learning algorithm is defined in Section \ref{s5}. Experimental results are shown in Section \ref{s6}, and Section \ref{s7} concludes the paper.

\section{Related Works}
\label{s2}
In recent years, there has been a large number of studies applying machine learning (ML) methods to wireless communications \cite{b12}. For example, in \cite{b5}, a DQN-based algorithms is proposed for power allocation in wireless networks. \cite{b8} proposed an correlated Q-learning algorithm to optimally allocate resources for network slicing. In \cite{b9}, the authors designed an algorithm to learn the optimal handover control strategy by using deep neural networks (DNNs). In \cite{b17}, a framework is defined to deploy artificial intelligence (AI) based algorithms in virtualized RANs. These algorithms only consider optimizing a single wireless network function and do not take into account the impact of other functions.

Furthermore, there are works that consider the joint allocation of several resources. In \cite{b4}, the authors proposed a reinforcement learning-based algorithm to jointly optimize power and radio resource allocation for ultra-reliable low latency communications in 5G. %In \cite{b10}, authors considered joint optimization of handover control and power allocation.
\cite{b10} jointly optimized the radio and cache resource allocation by transfer learning based method.
\cite{b13} studied joint power allocation and channel assignment for non-orthogonal multiple access network based on deep reinforcement learning and attention-based neural network. These papers have achieved good results in their respective contexts, but they are not applicable to conflict mitigation in O-RAN. In the O-RAN architecture, xAPPs are generally managed by different vendors. It maintains a higher degree of independence in the learning process with only necessary information sharing. On the other hand, above algorithms are often implemented by joint training or shared parameters. However, in the O-RAN architecture, different xAPPs may use heterogeneous frameworks and different learning parameters, which makes existing algorithms not applicable. 

\section{System Model}
\label{s3}
\begin{figure}[t]
\centerline{\includegraphics[width=3.1in]{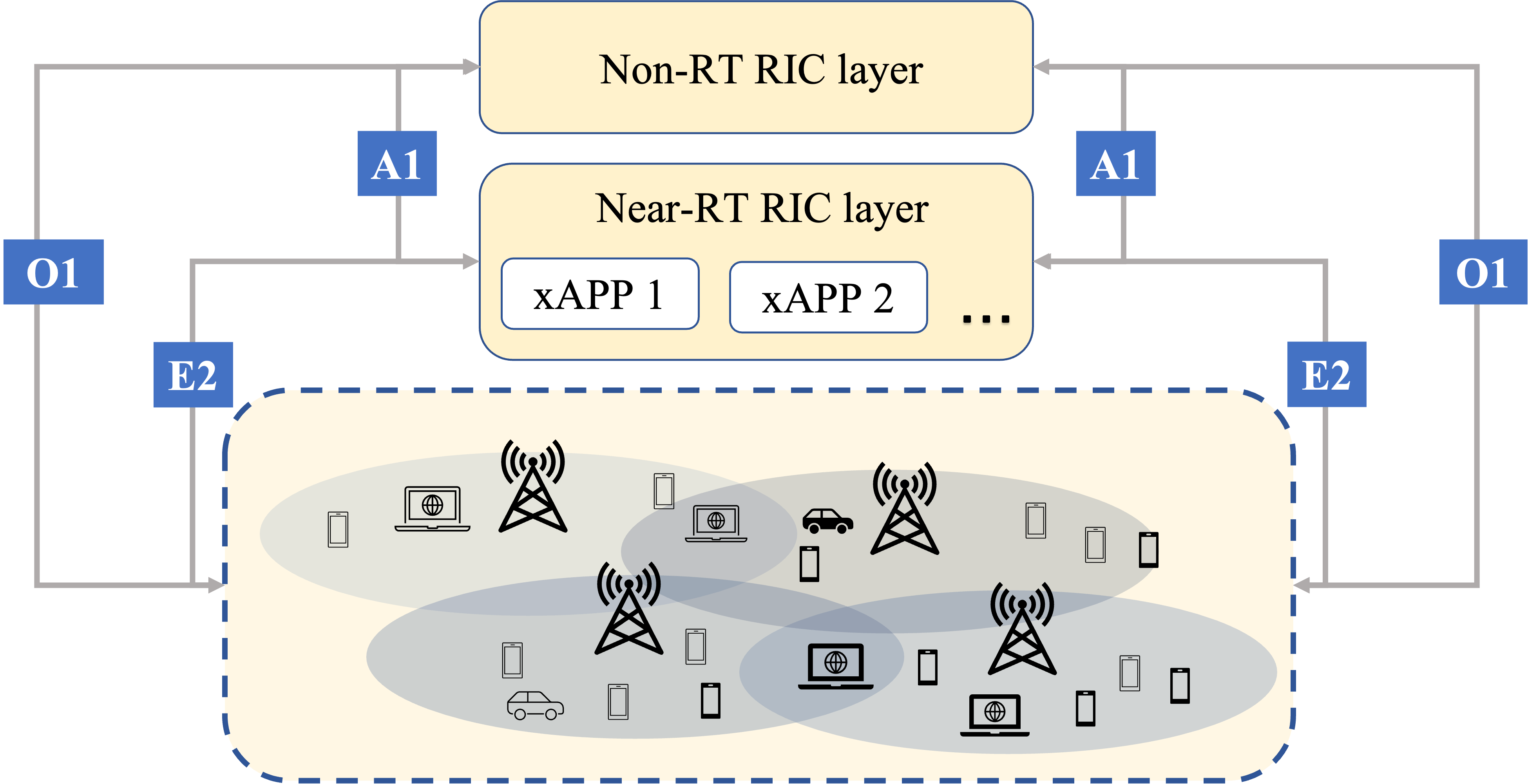}}
\caption{System model.}
\label{fig1}
\vspace{-15pt}
\end{figure}

In this paper, we consider a downlink orthogonal frequency-division multiplexing cellular system with $N$ base stations (BSs) serving $K$ users simultaneously. The considered wireless system architecture is shown in Fig.~\ref{fig1}. BSs are controlled by the near-RT RIC and non real-time (non-RT) RIC of O-RAN architecture. Radio controller of O-RAN architecture is divided into two parts, near-RT RIC and non real-time (non-RT) RIC. Near-RT RIC is used for RAN control and optimization. The role of the non-real-time layer is to provide guidance, as well as ML models to support near-RT RIC functions \cite{b14}. Near-RT RIC can communicate with BSs through interface E2 and non-RT RIC can communicate with BSs by interface O1. Interface A1 is used for communications between two RICs. 

%We assume users are automatically associated to the closest BS and each user updates its position
%\notered{do we need position info? where do we use it?} \notered{the user cell association should be signal strength based not location. In fact, here you are explaining the details of the simulator. all simulators use location implicity in channel models but in the end the decision is based on SINR. please also revise the optimization part. (4) should not be there.}\noteblue{Reply: The location is only used to decide whether a user is in the range of a base station. And the radio resource allocation is decided by the SINR with reinforcement learning algorithm. Not the location. We have removed (4) and descriptions here, and add descriptions in the simulation settings part.} %and association with BSs in each time slot. 
A resource block refers to the smallest unit of resources that can be allocated to a user and it is composed of 12 subcarriers. In our model, several consecutive resource blocks is bundled as a resource block group (RBG), and each BS has $M$ RBGs respectively. Here we consider the RBG as the smallest time-frequency resources to be allocated \cite{b4}. At time $t$, the list of users associated to BS $n$ can be described as $H_t^n$, where $\sum_{n\in N} |H_t^n| = K$. The transmission power allocated to the $m^{th}$ RBG of BS $n$ can be described as $P_t^{n,m}$.  

The signal interference noise ratio (SINR) between BS $n$ and user $k$ on RBG $m$ at time $t$ is described as $\eta_t^{n,m,k}$. It can be formulated as:
\begin{equation}
\eta^{n,m,k}=\frac{\alpha^{n,m,k} g^{n,k} P^{n,m}}{\sum_{n'\in N, n'\neq n}\sum_{k'\in K^{n'}} \alpha^{n',m,k'} g^{n',k} P^{n',m}+\sigma^2},\label{eq1}
 \end{equation}  
where $\alpha^{n,m,k}$ is a binary indicator that denotes whether BS $n$ allocates RBG $m$ to user $k$. $g^{n,k}$ denotes the channel coefficient between BS $n$ and user $k$. The transmission capacity of BS $n$ on RBG $m$ can be formulated as:

\begin{equation}
C^{n,m} = B_{m} log_{2}(1+\sum_{k\in K} \eta^{n,m,k}),\label{eq2}
\end{equation}
where $B_{m}$ denotes the bandwidth of RBG $m$. We assume user traffic follows Poisson arrivals with a mean arrival rate $\lambda$ and pending traffic is queued in a transmission buffer with limited buffer size. If the length of queuing data is longer than the buffer size, then the extra data will be discarded. The transmission rate of BS $n$ on RBG $m$ can be formulated as:

\begin{equation}
R^{n,m} =\begin{cases}
C^{n,m}, & C^{n,m}T < \sum_{k\in K}\alpha^{n,m,k}L^k\\
L^k/T, & C^{n,m}T >= \sum_{k\in K}\alpha^{n,m,k}L^k\\
\end{cases}
\label{eq3}
\end{equation}
%In \eqref{eq3}, 
where $T$ denotes the length of time slot and $L^k$ denotes the amount of remained data of user $k$ in the transmission buffer.

We consider user mobility in our system model. The user keeps a constant velocity and changes moving directions in each time slot with some probability. 
%In our model, users are moving at a fixed speed in random directions. %Their speed is denoted by $v_0$ and at time $t$, the angle of motion of user $k$ is denoted by $D_t^k$, which is given as:
%\begin{equation}
%\begin{split}
%&P(D_t^k = D_{t-1}^k) = p_0
%\\&P(D_t^k = random[0, 2\pi]) = 1-p_0
%\end{split}
%\label{eq3-a}
%\end{equation}
%where $p_0$ is the probability that the user maintains the current direction. 
%\noteblue{$p_0$ denotes the probability that the user maintains the current direction.}
%\begin{equation}
%\begin{split}
%&x_t^m = x_{t-1}^m + v_0\cos{D_t^k}
%\\ &y_t^m = y_{t-1}^m + v_0\sin{D_t^k}
%\end{split}
%\label{eq3-b}
%\end{equation}
 %In each time slot, each user has a probability of $p_0$ to keep the current moving direction and a probability of $(1 - p_0)$ to randomly choose a new direction. The nodes basically select a random point within a circle of radius R and move towards it
%\noteblue{In each time slot, we assume that each user basically keeps the current moving direction at possibility $p_0$ or selects a random point within a circle of radius $R$ and move towards it at possibility $(1-p_0)$}. 
%\noteblue{The user has a possibility of $P_0$ to change a random direction. Has a possibility of (1-$P_0$) to keep its direction. So I use "or", which means in a time slot, a user has two choices. We simplify this part }

Finally, the objective of the system is to maximize the total transmission rate for all BSs. It can be given as:
\begin{equation}
\begin{split}
 \underset{P^{m,n},\alpha^{n,m,k}}{max}\ &\Sigma_{n\in N}\Sigma_{m\in M}R^{n,m},\\
s.t.\ &(\ref{eq1})-(\ref{eq3}),\\& P_{min} \leq P^{m,n} \leq P_{max}, \forall n, m,\\
& \alpha^{n,m,k} = \{0,1\}, \forall n, m, k
\\& \Sigma _{k \in K}\alpha^{n,m,k} = 1, \forall n,m
\label{eq3-1}
\end{split}
\end{equation}
where $P_{min}$ and $P_{max}$ denote the maximum and minimum transmission power. %\noteblue{The variable 'g' in (1) is related to the distance between BS and users, and this is related to $x^m_t$ and $y_t^m$}

\section{Team Learning based xAPPs Management in O-RAN}
\label{s4}
\subsection{xAPP Interactions without Team Learning}
\begin{figure}[t]
\centerline{\includegraphics[width=3.1in]{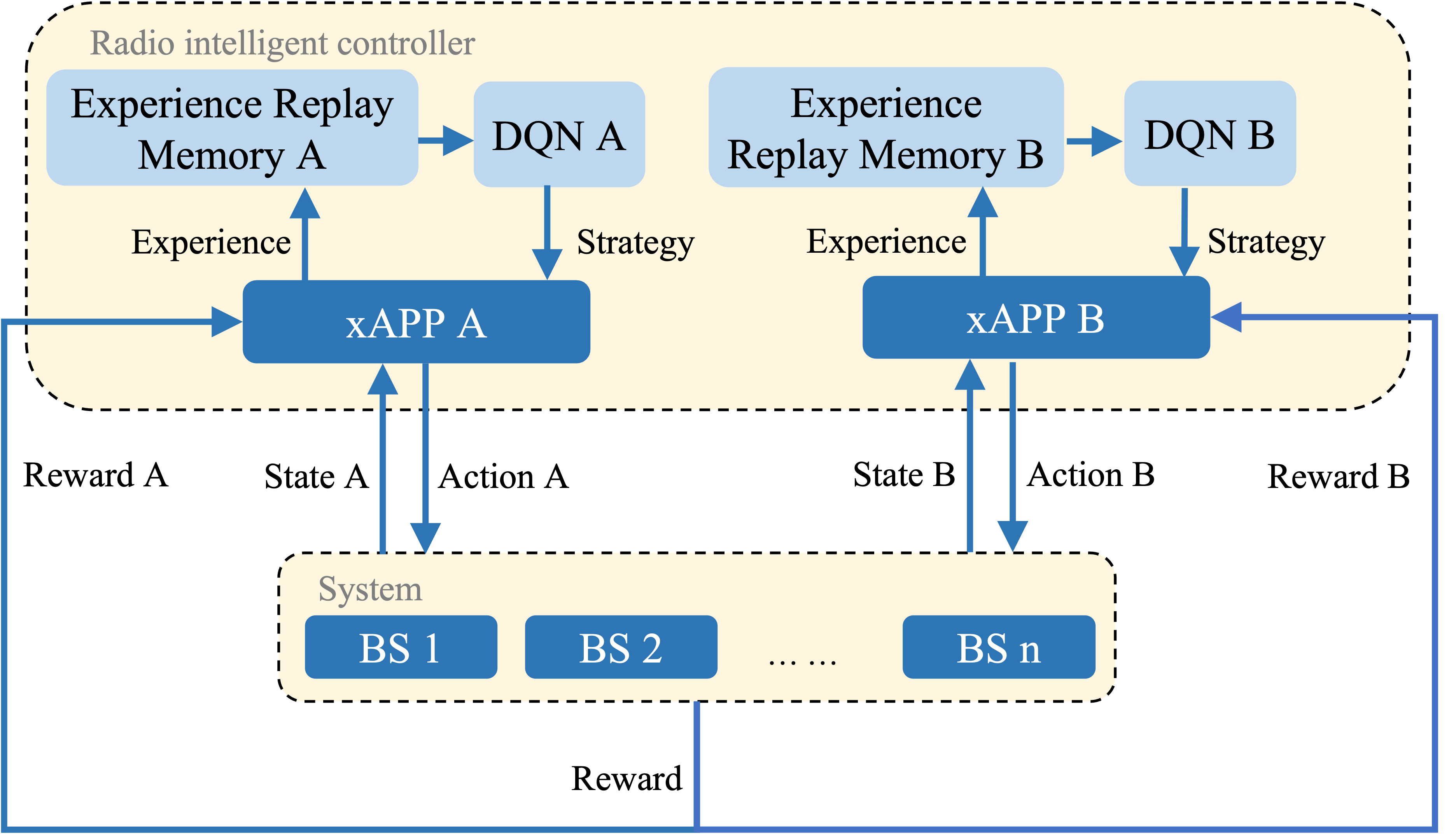}}
\caption{xAPP Interactions without Team Learning.}
\label{fig2}
\vspace{-15pt}
\end{figure}

When we deploy two intelligent xAPPs in the near-RT RIC layer of O-RAN architecture, conflicts may arise if they have overlapping objectives. In the absence of team learning, the system that contains two xAPPs is shown in Fig.~\ref{fig2}. Firstly, the intelligent xAPP A and B observe states from system and select actions accordingly. Then xAPP A and B simultaneously apply actions to the system, causing changes to the system that produces a reward and this reward is given back to xAPP A and B, respectively. The experience of xAPP A and xAPP B will be recorded in their own experience relay memory and used for DQN model training. Finally, the DQN module will generate strategies about action selection for two xAPPs.

In Fig.~\ref{fig2}, when an xAPP acts, it does not take into account the actions of other xAPPs. The selected action may be the optimal choice in its own view, but the performance of other xAPPs may be affected. For example, xAPP A may assign a high transmission power to a RBG. At the same time, xAPP B may reassign this RBG to user with a small traffic load. When xAPP A and xAPP B take actions together, the high transmission power does not bring high throughput as expected by xAPP A, but increases the power consumption and generates more interference. The goal of our team learning algorithm is to eliminate conflicts between actions from different team members.

\subsection{xAPP Interactions with Team Learning}
\begin{figure}[t]
\centerline{\includegraphics[width=3.1in]{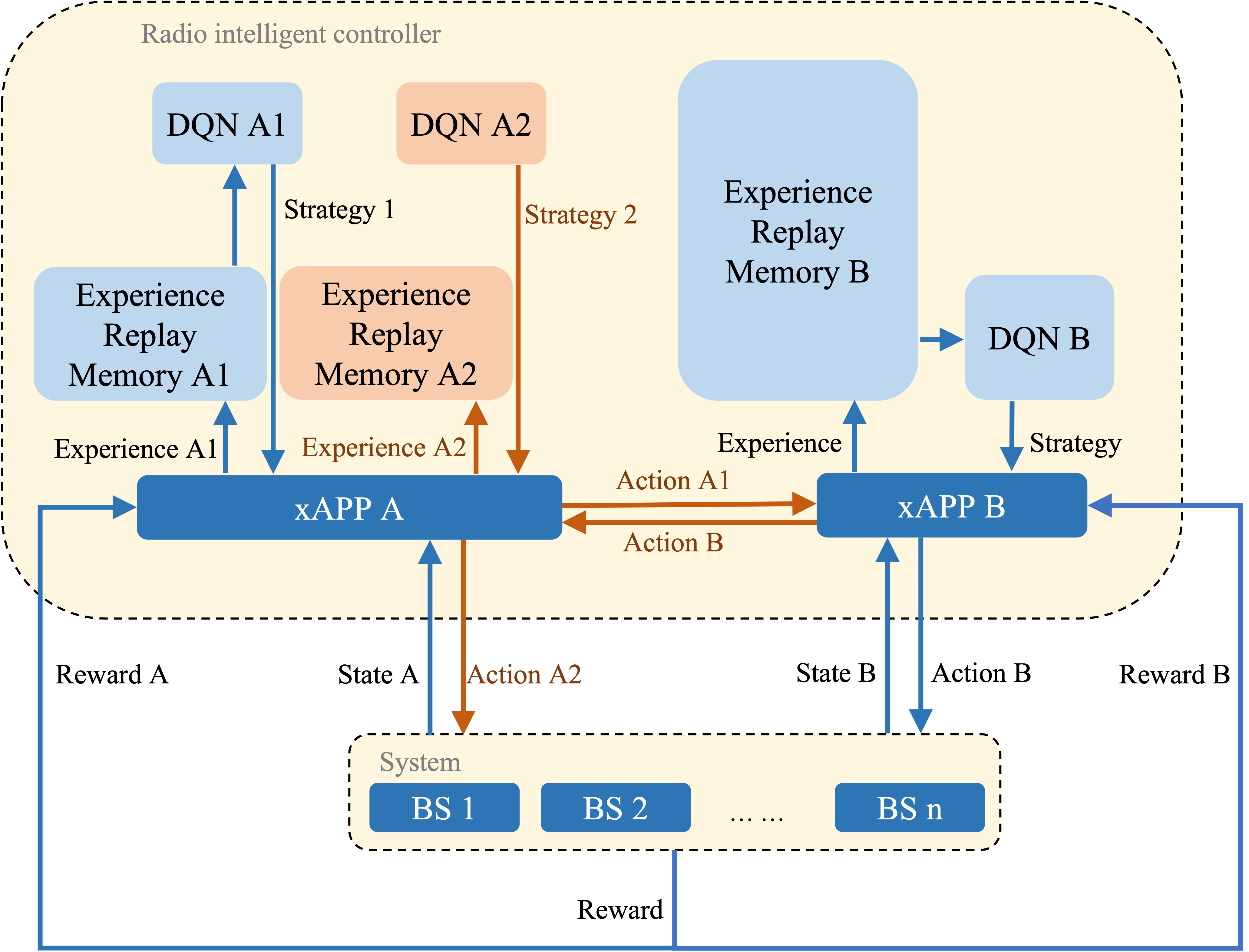}}
\caption{xAPPs Interactions with Team Learning.}
\label{fig3}
\vspace{-15pt}
\end{figure}

In the Fig.~\ref{fig2}, each xAPP only observes the state from the environment. However, in team learning based architecture, the xAPPs will include the action of other xAPPs in its own state, which means they can learn to avoid conflicts and cooperate better. In this paper, we propose an information exchange based team learning architecture, which is shown by Fig.~\ref{fig3}. Different than Fig.~\ref{fig2}, xAPP A keeps two experience replay memories and trains two DQNs. The first DQN is used to select a first-round action based on its own observation from environment. This action will be shared with xAPP B, and xAPP B selects its own action based on both the observation from environment and first-round action from xAPP A. Then the action of xAPP B will feed into xAPP A and will help xAPP A make a second-round action with the second DQN. 
%The experience stored in the first experience replay memory of xAPP A only includes the states observed from the environment. The experience in the second memory includes both the observation from the environment and the action of xAPP B. 
%which will be implemented in the actual system.

%In this way, xAPPs will take into account the actions that other xAPPs may take when selecting their own actions. With this information, they can make cooperation with each other and avoid possible conflicts.

\section{Team learning based xAPPs management algorithm}
\label{s5}
\subsection{Deep Q-learning}
In this paper, two xAPPs, power allocation and radio resource allocation, are both implemented by DQN. Q-learning is a popular value-based and model-free reinforcement learning algorithm. The core idea is that an agent uses a table of Q-values to represent the cumulative reward by taking a certain action $a$ under a certain state $s$ to maximize its expected long term reward. It can be formulated as:

\begin{equation}
Q(s,a) = E[r^t + \gamma Q(s^{t+1},a^{t+1})|s^t = s, a^t = a],\label{eq4}
\end{equation}
where $E[.]$ refers to the expectation operator. $r^{t}$ refers to the reward at time $t$. $\gamma$ is the discount coefficient. 

However, Q-learning suffers from long convergence time when faced with a huge state-action space, and consequently DQN is proposed. DQN uses Deep Neural Networks (DNNs) to estimate the Q-values and updates the parameters of DNNs with a stochastic gradient descent algorithm, which can be formulated as: 

\begin{equation}
\begin{split}
\theta^{t+1} &= \theta^t + \alpha[r^t + \gamma \mathop{max}\limits_{a'}Q(s^{t+1},a^{t+1};\theta^t) \\ & -Q(s^t,a^t;\theta^t)]\nabla Q(s^t,a^t;\theta^t),\label{eq6}
\end{split}
\end{equation}
where $\theta$ is the parameters of DQN. $\nabla Q(s^t,a^t;\theta^t)$ is the gradient. During the simulation, the current state, action, next state and reward are recorded as $(s^t, a^t, s^{t+1}, r^t)$. The recorded data is used for experience replay of DQN training.

\subsection{Power Allocation xAPP}\label{PA Model}
In this paper, we define two DQN models for power allocation. The first model is to determine the first-round action of power allocation, which is used to assist decision making of radio resource allocation. The state of this model is only related to the environment. The markov decision process (MDP) for the first DQN model of power allocation xAPP is given below:
\begin{itemize}
\item State: State includes transmission rate, transmission power, channel state information (CSI) and the length of queued data in the buffer \cite{b5}. The state of an agent is given as:
\begin{equation}
S_t^{m} = \{\Gamma_t^{n,m},R^{n,m}_t,p^{n,m}_t,L^{n,m}_t|n\in N\}\label{eq7}
\end{equation}
The $\Gamma^t_{n,m}$ is the logarithmic normalized CSI, which is given by:

\begin{equation}
\Gamma_t^{n,m} = \{log_2(1+\frac{\sum_{k \in K} \alpha_t^{n',m,k}g_t^{n',k}} {\sum_{k \in K} \alpha_t^{n,m,k}g_t^{n,k}})|n'\in N, n'\neq n\} \label{eq8}
\end{equation}
$L^{n,m}_t$ is the length of queued data in the buffer, which can be given as:

\begin{equation}
L^{n,m}_t = \Sigma _{k\in K} \alpha_t^{n,m,k}L^{k}\label{eq9}
\end{equation}

\item Action:
In our model, we divide the transmission power into $B$ levels according to the maximum %transmission power 
and the minimum transmission power. The action is to choose a power level for each RBG of each BS, which is given as:

\begin{equation}
\begin{split}
a^{n,m}_t &= \{P_{min}, P_{min}+ \frac{P_{max}-P_{min}}{B-1},...,P_{max}\}
\\
A_t^m &= \{a^{n,m}_t|n\in N\}
\end{split}
\label{eq10}
\end{equation}

\item Reward:
The reward is defined as the  total throughput since our objective is to improve throughput:
\begin{equation}
r_t = \Sigma _{m\in M}\Sigma _{n\in N} R^{n,m}_t\label{eq11}
\end{equation}
Where $R_t^{m,n}$ is the transmission rate of BS $n$ on RBG $m$.
\end{itemize}

The second model is used to determine the power allocation action for the second-round. This action is the actual action to be performed by power allocation. The state of this model is related to the environment and the information passed by the radio resource allocation. The MDP for the second DQN model of power allocation is  given  below:
\begin{itemize}
\item State: The sate is given as:
\begin{equation}
\begin{split}
S_t^{m} = \{&\Gamma_t^{n,m'},R^{n,m'}_t,p^{n,m'}_t,\\&\Gamma_t^{n',m'},R^{n',m'}_t,p^{n',m'}_t,L^{n',m'}_t|n\in N\}\label{eq15}
\end{split}
\end{equation}
%Here 
where $m'$, $n'$ is given as:
\begin{equation}
\begin{split}
&m',n' \in \\ & \{m'\in M, n' \in N, {\alpha_{t+1}^{n',m',k}} =1|k \in \{\alpha_t^{n,m,k} =1 \}\}\label{eq16}
\end{split}
\end{equation}
$\alpha_{t+1}^{n',m',k}$ is obtained from the intended action of radio resource allocation model.

\item Action and Reward: The definition of action and reward of the second model is the same as the first model.

\end{itemize}

The first DQN is used to select the intended action which will be informed to xAPP radio resource allocation. The second DQN is used to select the actual action.

\subsection{Radio Resource Allocation xAPP}\label{RA Model}
In radio resource allocation, each BS selects and distributes resource blocks among its own associated users, and we define a BS as an agent. We only define one DQN for radio resource allocation because it only needs one round action. The state of this model is a combination of observation from environment and information from power allocation. The MDP is defined by:

\begin{itemize}
\item State: %Same as power allocation, 
the state of radio resource allocation agent includes transmission rate, transmission power, CSI and the length of queuing data in the buffer. The state is given as:

\begin{equation}
\begin{split}
S_t^{n} = \{\alpha_t^{n,m,k}\Gamma_t^{n,m},\alpha_t^{n,m,k}R^{n,m}_t, \alpha_t^{n,m,k}p^{n,m}_t,\\ \alpha_t^{n,m,k}L^{n,m}_t, p^{n,m}_{t+1},|m\in M, k \in K \},\label{eq17}
\end{split}
\end{equation}
where $p^{n,m}_{t+1}$ is obtained from the intended action of power allocation model. Since the radio resource allocation policy of a BS is only related with its associated users, we only consider the features related to these users. So we add indicator $\alpha_t^{n,m,k}$ in the state.

\item Action:
The action of BS $n$ is to choose a user by: % from the first $D$ users in the list of :
\begin{equation}
\begin{split}
a^{n,m}_t &= \{k_0, k_1, ..., k_{D-1}| k_d \in H^n_t\}
\\
A_t^n &= \{a^{n,m}_t|m\in M\}
\end{split}
\label{eq13}
\end{equation}

\item Reward:
The reward of radio resource allocation is the total throughput of each base station, which is given as:
\begin{equation}
r_t = \Sigma _{m\in M}\Sigma _{n\in N} R^{n,m}_t\label{eq14}
\end{equation}
\end{itemize}

\subsection{Implementation of team learning in O-RAN architecture}
Our proposed scheme of xApp coordination can be implemented as an overlay either in the RT RIC or non-RT RIC, or split between these two. According to the O-RAN architecture, model training can be instantiated in the non-RT RIC layer by AI servers such as Acumos AI \cite{b18} while trained models are deployed in the near-RT RIC layer via containerized images to handle action selection \cite{b19}. The RICs and O-CU/O-DU communicate by bi-directional open interfaces (A1 and O1, respectively). In particular, in this work, non-RT RIC layer can obtain the information of power and radio resource allocation xAPPs and train their DQN models by AI servers. Then the trained models can be used by the xAPPs in the non-RT RIC layer for action selection.  

\section{Numeric Results}
\label{s6}
\subsection{Simulation settings}
 %The large-scale fading of wireless channel is modeled according to the LTE standard and the path loss can be formulated as: $Path Loss(dB) = 120.9+37.6 log10(d)+10 log10(z) (dB)$. Here $z$ is a log-normal random variable with 8 dB standard deviation. $d$ is the distance between user and BS. The maximum transmission power $P_max$ and minimum transmission power $P_min$ are respectively 38 dBm and 1 dBm. The additive white Gaussian noise (AWGN)\noteblue{ do not define an abbreviation unless you use it frequently} is -114 dBm. For the carrier configuration, we use an 20 MHz bandwidth carrier at 4GHz and it is divided into 100 resource blocks. Each resource block has 12 subcarriers with 15kHz subcarrier spacing. Then we decide 100 resource blocks into $M = 12$ RBGs. The first 11 RBGs respectively have 8 resource blocks and the last RBG has 12 resource blocks. Considering the near RT RIC layer operates in the time frame between 10ms and 1s, the time slot of simulation is set as 100ms.\noteblue{delete}

Table.\ref{table1} includes networks settings of our simulations. We use Python simulation platform and the reinforcement learning and team learning algorithms are based on the deep-learning package TensorFlow. The algorithm is simulated with 4 BSs and 30 users. The amount of data generated by each user per time slot follows a Poisson distribution with the central value of the distribution varying from 3 Mbps to 6 Mbps. 
The users move with a constant speed during the simulation, and the probability of changing direction is 0.3 in each time slot.  %The direction of their movement is zero initialized, and for each user, the probability of keeping current moving direction at each time slot is 0.7. 

We use two four-layer neural networks in the DQN for power allocation and radio resource allocation. According to the dimension of states and actions, we set the neuron numbers of two hidden layers of power allocation model as 256 and 128 and the neuron numbers of radio resource allocation model as 512 and 256. The discount factor is set as 0.2. The initial learning rate of both power allocation and radio resource allocation is set as 0.001. We run 20000 time slots in each experiments and apply an adaptive $\epsilon$-greedy learning strategy with an initial exploration rate of 0.3 \cite{b6}. 

We compared our proposed team deep Q-learning (TDL) algorithm with independent deep Q-learning (IDL) algorithm, which is shown in Fig.~\ref{fig2}. In IDL, xAPPs only observe information from the environment and do not communicate with each other. We change the traffic load of each user between 3-6 Mbps and the user moving speed between 0-30 m/s and compare the performance of TDL and IDL in different scenarios.

\begin{table}[t]
\caption{Simulation settings}
\label{table1}
\begin{tabular}{ll}
\hline
Parameter              & Value                                                                                                                                                                                                             \\ \hline
Networking Environment & \begin{tabular}[c]{@{}l@{}}4 Base Stations with 1 kilometre inter-site \\ distance, 30 users.\end{tabular}                                                                                                         \\ \hline
Propagation            & $\beta$ = 120.9+37.6 log10(d)+10 log10(z) dB                                                                                                                                                                            \\ \hline
Carrier configuration  & \begin{tabular}[c]{@{}l@{}}20MHz bandwidth, 100 resource blocks, \\ 12 subcarriers per resource block, 12 RBGs.\end{tabular}                                                                                    \\ \hline
PHY configuration      & \begin{tabular}[c]{@{}l@{}}Maximum transmission power of 38 dBm, \\ minimum transmission power of 1dBm\\ Additive white Gaussian noise = -114dBm.\end{tabular}                                                     \\ \hline
%Movement pattern       & \begin{tabular}[c]{@{}l@{}}Users are initially randomly distributed. \\ Random wandering at a fixed velocity.\end{tabular}                                                                                        \\ \hline
Traffic model          & \begin{tabular}[c]{@{}l@{}}Poisson distribution  with varying load \\between 3-6 Mbps.\end{tabular} \\ \hline
Simulation time        & 20,000 time slots. 1 time slot is 100ms.                                                                                     \\ \hline
\end{tabular}
\end{table}

\begin{figure}[!b]
\centerline{\includegraphics[width=3.1in]{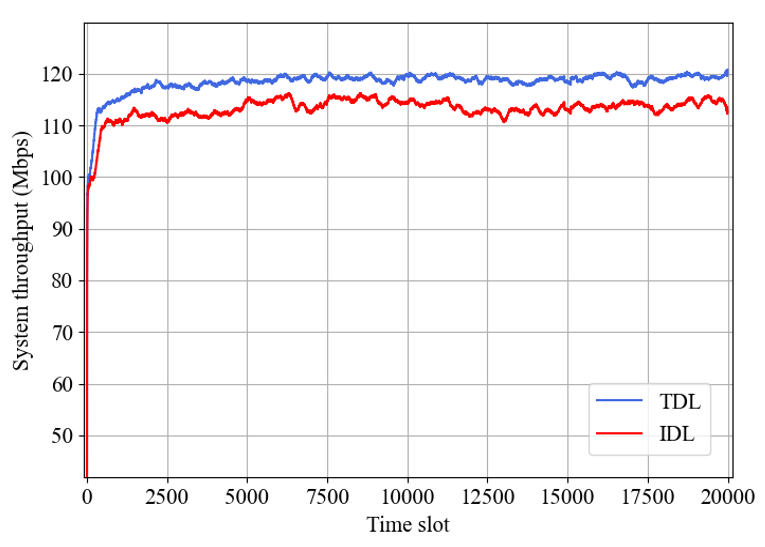}}
\caption {System throughput (traffic load = 4 Mbps, user speed = 20 m/s).}
\label{fig4}
\end{figure}

\begin{figure}[!b]
\centerline{\includegraphics[width=3.1in]{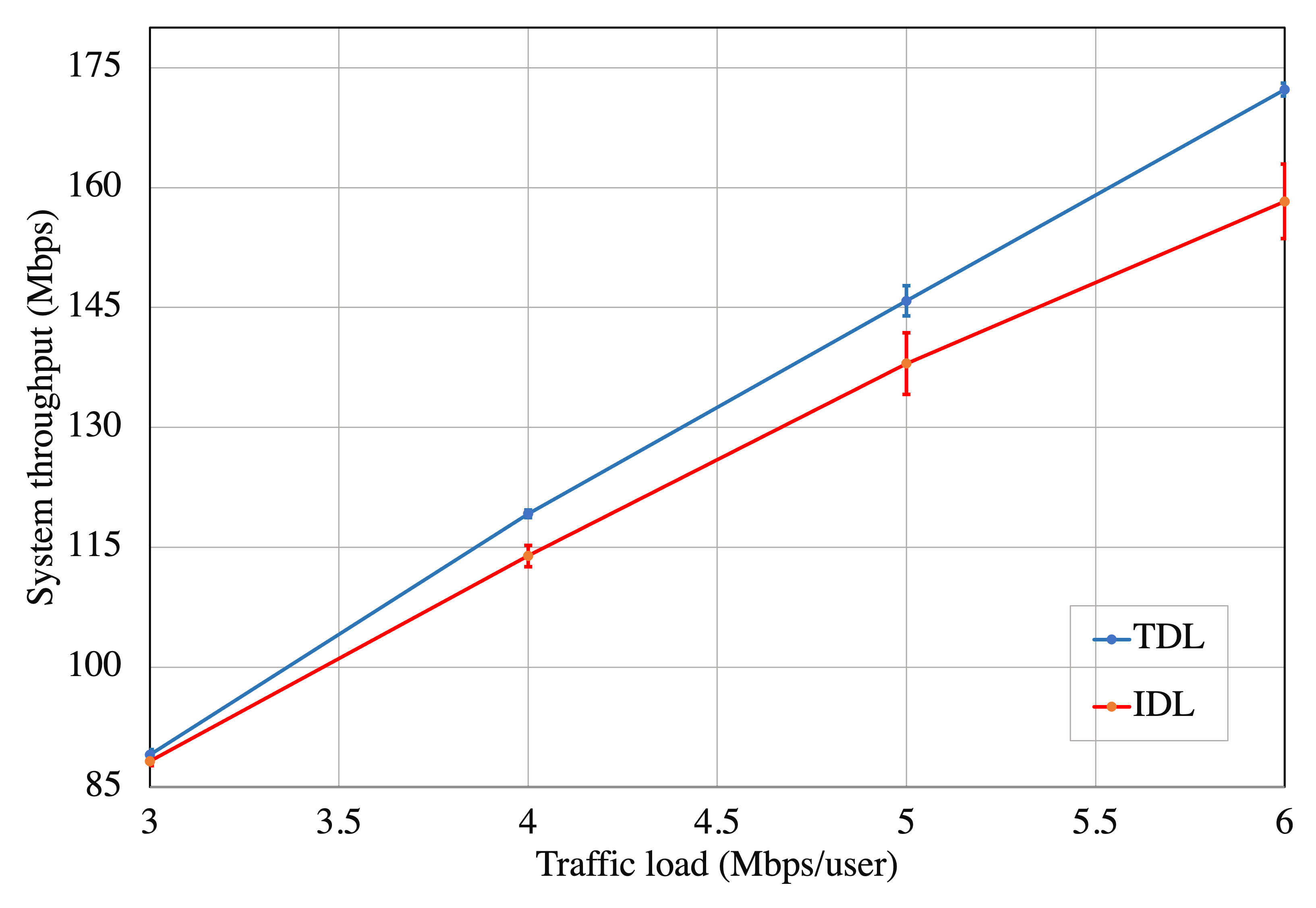}}
\caption {System throughput with different traffic load (user speed = 20 m/s).}
\label{fig5}
\end{figure}

\begin{figure}[!t]
\centerline{\includegraphics[width=3.1in]{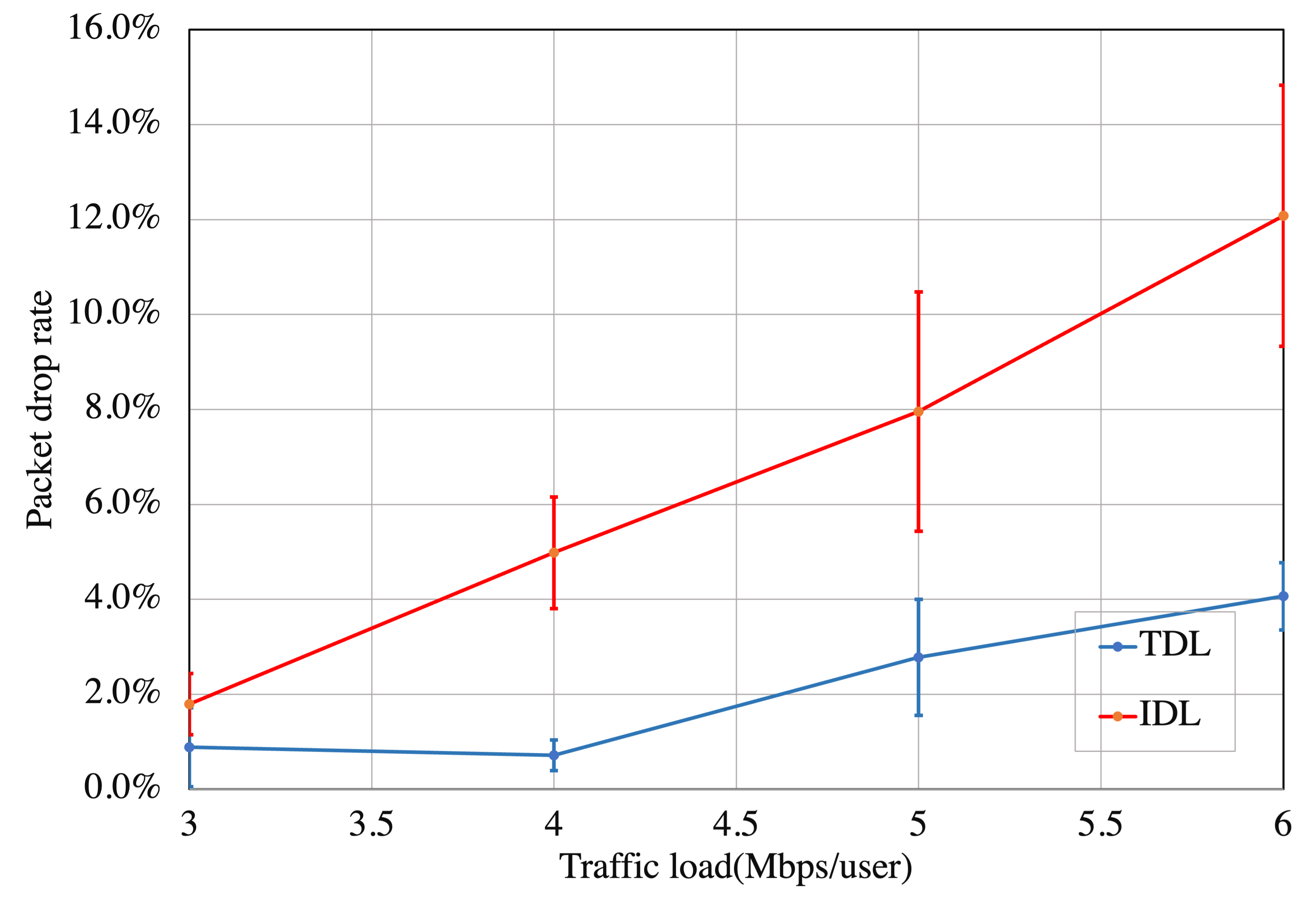}}
\caption {PDR with different traffic load (user speed = 20 m/s).}
\label{fig6}
\end{figure}

\begin{figure}[!t]
\centerline{\includegraphics[width=3.1in]{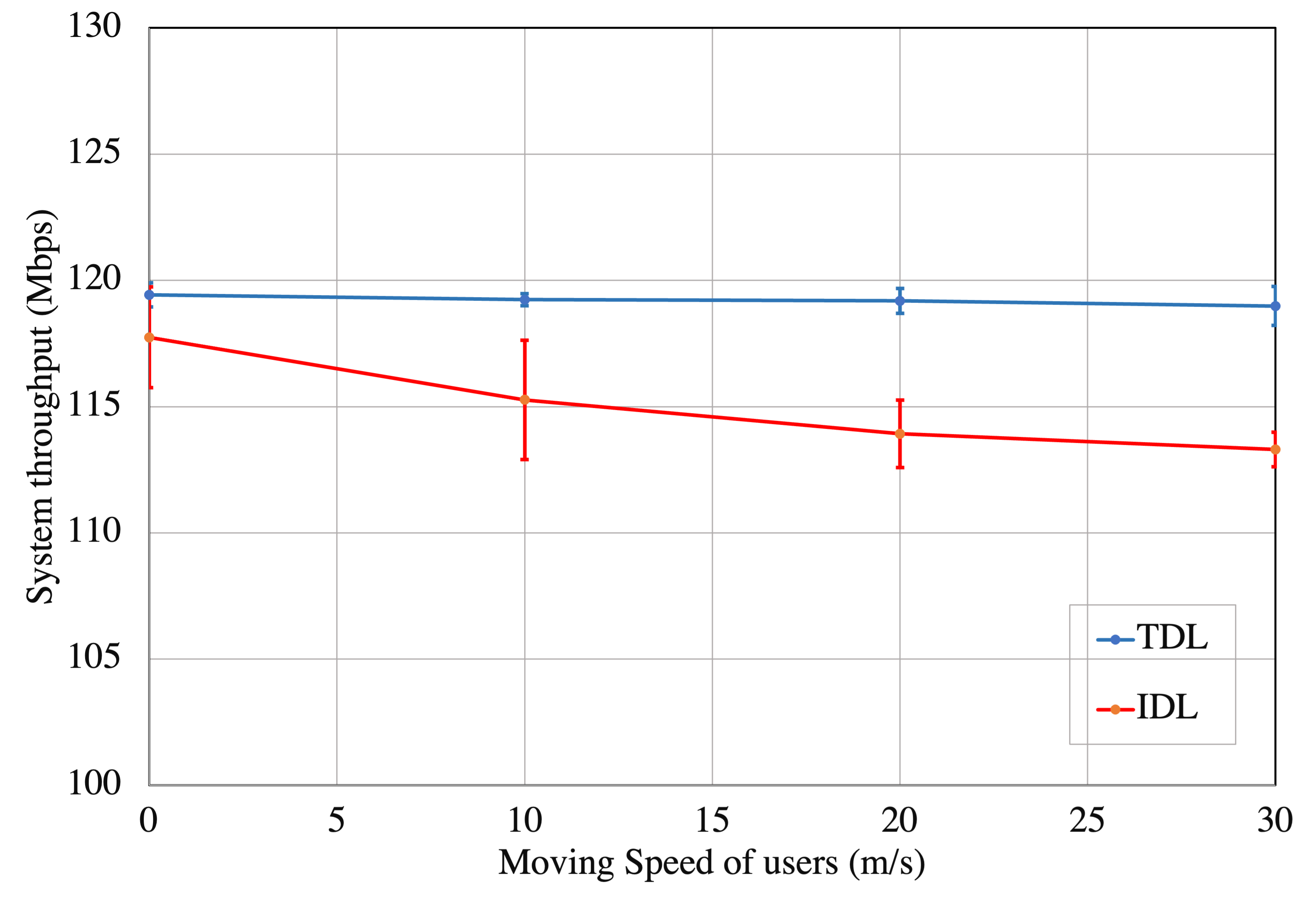}}
\caption {System throughput with different user moving speed (traffic load = 4 Mbps).}
\label{fig7}
\vspace{-10pt}
\end{figure}

\subsection{Simulation Results}
In this section, we compare the convergence as well as the throughput of all BSs and the packet drop rate (PDR) of the system for TDL and IDL. 

Fig.~\ref{fig4} shows the convergence curves of TDL and IDL during the learning process and compares the system throughput with TDL and IDL at 4 Mbps traffic load when the speed of users is 20 m/s. We can observe from Fig.~\ref{fig4} that the throughput with TDL is more stable and higher than that with IDL during the whole simulation process. When the simulation performance is stabilized, the system can achieve a 4.6\% higher throughput with TDL. 

Fig.~\ref{fig5} and Fig.~\ref{fig6} show the system throughput and PDR when the speed of users is 20 m/s and the average traffic load of users changes from 3 Mbps to 6 Mbps. In Fig.~\ref{fig5}, we observe that the system throughput with TDL is higher than that with IDL. Correspondingly, we can also see from Fig.~\ref{fig6} that the PDR with TDL is lower than that with IDL. The gap between them becomes more noticeable when the traffic load becomes higher. This is because when the traffic load is relatively low, less data needs to be transmitted, and even a poorer allocation policy can handle the transmission task. When the traffic load is high there is more data to be transmitted, hence the advantage of a good allocation policy is more significant. When the traffic load is 6 Mbps, the TDL achieves 8.8\% higher throughput and 64.8\% lower PDR than IDL.

Fig.~\ref{fig7} shows the system throughput when the average traffic load of users is 4 Mbps and the speed of users changes from 0 m/s to 30 m/s. Fig.~\ref{fig7} presents that the system throughput of TDL is higher than that IDL especially when the user speed is high. This is because in the case of fast moving users, the allocation strategies will change more frequently, and there will be more conflicts between xAPPs. As a result, cooperation becomes more important. Note that, the mobility of users are limited in a circular region therefore mobility doesn't cause handover in our experiments. When the moving speed of users is 30 m/s, the TDL achieves 5.0\% higher throughput than IDL.

The above results show that for different traffic loads and user speeds, TDL results in higher throughput and lower PDR compared to IDL.

\section{Conclusion}
\label{s7}
In this work, we proposed a team learning algorithm to mitigate conflicts between xAPPs in the O-RAN architecture. This scheme is also applicable to other multi-vendor RANs of the future. The core idea is to make xAPPs share action information they intend to take. Then the intended actions of other xAPPs will be used as a part of the state for DQN training and the selection of actions. We used two xAPPs, power allocation and radio resource allocation, as examples to explain how the proposed team learning algorithm is specifically implemented and applied to the O-RAN architecture. Simulation results show that our proposed team learning algorithm can achieve higher system throughput and lower PDR compared with cases where team learning is not applied. In the future, we will consider other xAPPs and extend our algorithm to use cases with more than two xAPPs.

\section*{Acknowledgement}
This work is funded by Canada Research Chairs programs and NSERC Collaborative Research and Training Experience Program (CREATE) under Grant 497981.

\end{document}